\documentclass[conference]{IEEEtran}
\IEEEoverridecommandlockouts
\usepackage{cite}
\usepackage{amsmath,amssymb,amsfonts}
\usepackage{algorithmic}
\usepackage{graphicx}
\usepackage{textcomp}
\usepackage{xcolor}
\usepackage{multirow}
\usepackage{tabularx}
\usepackage{url}
\usepackage{subfig}

\def\BibTeX{{\rm B\kern-.05em{\sc i\kern-.025em b}\kern-.08em
    T\kern-.1667em\lower.7ex\hbox{E}\kern-.125emX}}
\begin{document}


\title{Benchmarking and Dissecting the Nvidia Hopper GPU Architecture}

\author{Weile Luo$^{1}$, Ruibo Fan$^{1}$, Zeyu Li$^{1}$, Dayou Du$^{1}$, Qiang Wang$^{2,\dagger}$, Xiaowen Chu$^{1,3,\dagger}$ 
\thanks{$^{1}$The Hong Kong University of Science and Technology (Guangzhou), Guangzhou, China, {\tt\small \{wluo976, rfan404, zli755, dda487\}@connect.hkust-gz.edu.cn, xwchu@ust.hk}}
\thanks{$^{2}$School of Computer Science and Technology, Harbin Institute of Technology (Shenzhen), Shenzhen, China, {\tt\small qiang.wang@hit.edu.cn}}
\thanks{$^{3}$The Hong Kong University of Science and Technology, Hong Kong SAR, China.}
\thanks{$^{\dagger}$Corresponding authors.}
}

\maketitle

\begin{abstract}
Graphics processing units (GPUs) are continually evolving to cater to the computational demands of contemporary general-purpose workloads, particularly those driven by artificial intelligence (AI) utilizing deep learning techniques. A substantial body of studies have been dedicated to dissecting the microarchitectural metrics characterizing diverse GPU generations, which helps researchers understand the hardware details and leverage them to optimize the GPU programs. However, the latest Hopper GPUs present a set of novel attributes, including new tensor cores supporting FP8, DPX, and distributed shared memory. Their details still remain mysterious in terms of performance and operational characteristics. In this research, we propose an extensive benchmarking study focused on the Hopper GPU. The objective is to unveil its microarchitectural intricacies through an examination of the new instruction-set architecture (ISA) of Nvidia GPUs and the utilization of new CUDA APIs. Our approach involves two main aspects. Firstly, we conduct conventional latency and throughput comparison benchmarks across the three most recent GPU architectures, namely Hopper, Ada, and Ampere. Secondly, we delve into a comprehensive discussion and benchmarking of the latest Hopper features, encompassing the Hopper DPX dynamic programming (DP) instruction set, distributed shared memory, and the availability of FP8 tensor cores. The microbenchmarking results we present offer a deeper understanding of the novel GPU AI function units and programming features introduced by the Hopper architecture. This newfound understanding is expected to greatly facilitate software optimization and modeling efforts for GPU architectures. To the best of our knowledge, this study makes the first attempt to demystify the tensor core performance and programming instruction sets unique to Hopper GPUs.
\end{abstract}

\begin{IEEEkeywords}
Instruction Latency, Tensor Core, PTX, Hopper, DPX, Asynchronous Execution, Distributed Shared Memory
\end{IEEEkeywords}

\section{Introduction}
Graphics Processing Units (GPUs) have experienced a significant leap in their capacity to accelerate a wide array of applications, spanning from neural networks to scientific computing. This growth has been particularly propelled by the emergence of large language models (LLMs), where models like GPT-3, boasting over 150 billion parameters, stand as prime examples \cite{floridi2020gpt}. Modern GPU architectures, such as Ampere, Ada, and Hopper, embody cutting-edge features like tensor cores and high-bandwidth memory, meticulously crafted to elevate artificial intelligence applications. These GPUs have now firmly established themselves as the bedrock of computing infrastructure in high-performance clusters.

Nvidia consistently introduces new GPU architectures every two years, incorporating advanced features. However, detailed micro-architecture information about these features is often limited, making precise quantification challenging. In-depth studies are increasingly essential to understand the impact of these advancements on application performance.
The tensor core (TC) unit was initially introduced with the Volta architecture, focusing on accelerating deep neural networks with FP16 and FP32 precision operations. Subsequent Ampere architectures expanded TC capabilities to include sparsity and a broader range of data precisions such as INT8, INT4, FP64, BF16, and TF32. The Hopper architecture extended this further, introducing support for FP8 precision, significantly enhancing LLM training and inference acceleration.
While a recent study \cite{sun_tpds_2023} discussed TC programmability on Hopper, assembly code analysis and microbenchmarks were still conducted on Ampere and Turing, highlighting the need for further research specifically on Hopper tensor cores.

\begin{figure}[htbp]
    \centering
    \includegraphics[width=0.96\linewidth]{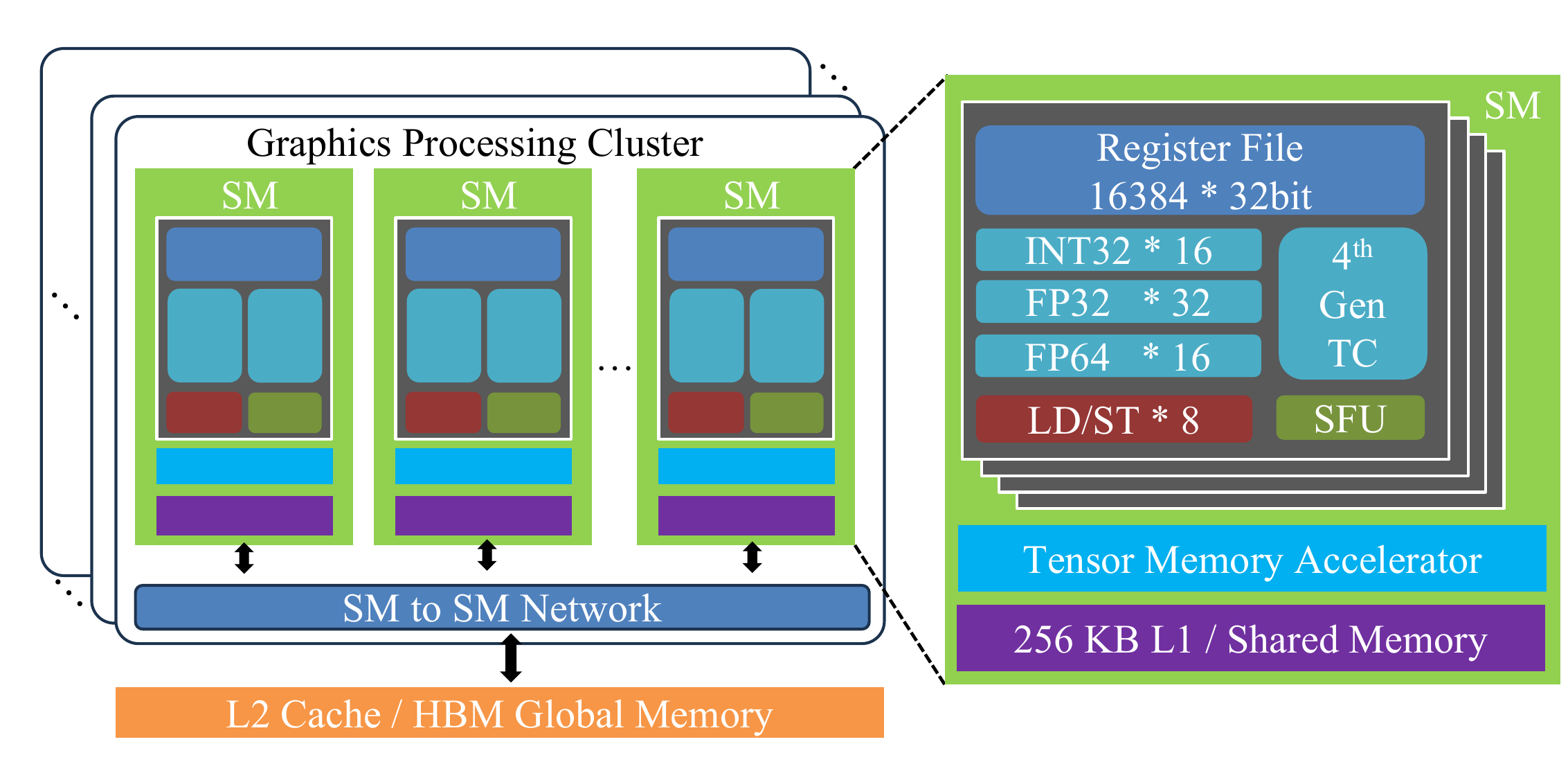} 
    \caption{Hopper architecture}
    \label{fig:hopper}
    \vspace{-1.0 em}
\end{figure}

In addition to the new tensor core, as shown in Fig. \ref{fig:hopper}, Hopper introduces innovative features: Dynamic Programming X (DPX) instructions, distributed shared memory (DSM), and an enhanced asynchronous execution mechanism (Tensor Memory Accelerator) for diverse scenarios.
DPX instructions accelerate a wide range of dynamic programming algorithms, often involving numerous minimum/maximum operations for comparing previously computed solutions. DSM enables direct SM-to-SM communications, including loads, stores, and atomics across multiple SM shared memory blocks.
Hopper supports asynchronous copies between thread blocks within a cluster, enhancing efficiency. However, detailed implementation and performance specifics remain undisclosed in existing literature. Unveiling these technical details is crucial for programmers to optimize AI applications effectively and leverage the new features of modern GPUs.

In this study, we conduct a comprehensive benchmarking of the latest GPU architectures (Ampere, Ada, and Hopper), focusing on key features like tensor cores and asynchronous operations. To the best of our knowledge, our research presents a pioneering analysis of the new programming interfaces specific to the Hopper architecture, offering a unique horizontal performance comparison among these cutting-edge GPU architectures. Many of our findings are novel and being published for the first time, providing valuable insights.
We highlight the contributions of our work as follows. 
\begin{itemize}
    \item We conduct detailed instruction-level testing and analysis on memory architecture and tensor cores across three GPU generations with different architectures. Our analysis highlights the unique advantages and potential of the Hopper architecture.
    \item We compare AI performance across recent GPU generations, examining latency and throughput of tensor cores at the instruction level, transformer engines at the library level, and real LLM generation at the application level. This comprehensive analysis provides valuable insights for AI system and application design and optimization, aiding informed decisions by researchers, developers, and manufacturers for next-generation AI solutions.
    \item Our research represents the inaugural exploration of Hopper architecture's distinctive features, encompassing DPX, asynchronous memory operations, and distributed shared memory. These innovations hold significant potential for enhancing GPU programming methodologies. Our study, by evaluating the performance of these features, contributes to performance modeling and algorithm design in dynamic programming and scientific computing. This contribution may fully unlock the GPU performance potential, driving advancements in the field.
\end{itemize}
\section{Related Work}
Analyzing GPU microarchitectures and instruction-level performance is crucial for modeling GPU performance and power \cite{isca2010_hong,taco2021_portable,hpca2019_hybrid,tpds2020_gpudvfs,mei2017survey,arafa2020verified,van2022isolating,arafa2019low}, creating GPU simulators \cite{accelsim_isca2020,bakhoda2009analyzing,gpuwattch_isca2013}, and optimizing GPU applications \cite{bakhoda2009analyzing,ho2017exploiting,yan_optimize_ppopp2020}. Early research \cite{wong2010demystifying,mei2014benchmarking,mei_tpds_2017,jia2018dissecting,jia2019dissecting} extensively dissected undisclosed GPU microarchitecture characteristics, particularly in older architectures like Fermi, Kepler and Maxwell. Jia et al. \cite{jia2018dissecting,jia2019dissecting} further evaluated the Volta and Turing GPU architectures, adding some special measurements for the new tensor cores. 

Fused Matrix Multiplication Accumulation (MMA), a critical operation in AI, is predominantly accelerated by tensor cores (TCs) in Nvidia GPUs since the Volta architecture. To harness TCs' full potential, extensive research has dissected TCs across Volta, Turing, and Ampere architectures, focusing on compute throughput and register mapping.
Early studies \cite{markidis_tensor_ipdpsw2018,benchmark_Martineau_eupar2019} examined the first-generation tensor cores on Volta GPUs, emphasizing legacy \emph{wmma} APIs and benchmarks using vendor CUBLAS and CUTLASS libraries. Subsequent work by Jia et al. \cite{jia2018dissecting,jia2019dissecting} expanded this analysis to the Turing architecture, providing preliminary assembly code analysis of \emph{wmma} APIs for TCs. However, comprehensive instruction-level microbenchmarks to fully exploit TC performance and numeric behaviors were lacking.
Yan et al. and Md Aamir et al. \cite{yan_ipdps_2020,yan_optimize_ppopp2020,raihan_ispass2019} delved into assembly code (SASS level) benchmarking, demystifying Volta and Turing TCs. Their focus was on assembly-level optimizations for matrix multiplication performance. Additionally, they optimized half-precision matrix multiplication on TCs.
A study by Fasi et al. \cite{fasi2021numerical} scrutinized the numeric behaviors of TCs, exploring rounding modes and subnormal behaviors of TF32, BF16, and FP16 data types. However, it's noted that \emph{wmma} APIs have limitations in operand shapes and cannot fully leverage the new sparse matrix multiplication features on Ampere and Hopper GPUs.

On the contrary, the new \emph{mma} programming interface was proposed since the Turing architecture and has evolved to support sparse matrix multiplication (\emph{mma.sp}) on the architecture above Ampere. 
Sun et al. \cite{sun_tpds_2023} conducted a comprehensive study of instruction-level microbenchmarks on the current \emph{mma} APIs (\emph{ldmatrix, mma and mma.sp}) instead of legacy \emph{wmma} APIs, exploring the full performance of tensor cores, the computation numeric behaviors of low-precision floating points, and the new sparse matrix multiplication feature of Turing and Ampere GPUs. 
As for the newest Hopper, the SASS codes of \emph{wgmma} and \emph{mma} can be even more diverse to support a wide range of computational precisions. The usage as well as their performance is still uncovered.

In addition to performance, energy efficiency is also an often discussed factor. In addition to the above literature that models GPU power, some work also benchmarks the application level. \cite{tang2019impact} empirically investigated the impact of GPU Dynamic Voltage and Frequency Scaling (DVFS) on energy consumption and performance during deep learning, testing various GPU architectures, DVFS settings, and DNN configurations. \cite{wang2020benchmarking} performed an empirical comparison of performance and energy efficiency across different AI accelerators from multiple vendors when training DNNs.

Despite the popularity of tensor cores and AI, another trend is the support of DPX, asynchronous operation support and distributed shared memory. 
To this end, benchmarking and dissecting the performance details of the modern GPU architectures is necessary and emerging. Revealing them helps programmers investigate more optimization opportunities.
\section{Methodology}
\subsection{Memory Unit Access Latency and Throughput}
In this subsection, we focus on two memory performance metrics: latency and throughput. Our memory test in this subsection uses a method similar to P-chase microbenchmark, which was first introduced on \cite{Saavedra_Smith_1995}\cite{Saavedra_Barrera_1992}. Below we will introduce how we test these two metrics. 
\subsubsection{L1 Cache}
For the latency test, we first load the data from global memory to the L1 cache using the $ca$ modifier. Then we use a thread to access this L1 cache to obtain the latency.
For the throughput test, we also first load the memory into the L1 cache using the $ca$ modifier. Since the L1 cache is exclusive to SM, we only issue a block with 1024 threads to repeatedly access the L1 cache. We record the time consumed and the amount of data accessed to calculate the bandwidth of the L1 cache.
\subsubsection{Shared memory}
Testing the shared memory is similar to testing the L1 cache. The only difference is that there is no need to specify modifiers to explicitly warm up shared memory. We can test it directly by declaring shared memory. Since shared memory can only be accessed within a block (distributed shared memory is not considered in this subsection), we use one thread to test latency and a block with 1024 threads to test bandwidth just like testing the L1 cache.
    
\subsubsection{L2 Cache}
For the latency test, we use the same method as for L1 cache testing. The only difference is that the $cg$ modifier is used instead of $ca$, ensuring that the cache we load is L2.
For the throughput test, we first load the memory into the L2 cache using the $cg$ modifier. Since the L2 cache is shared by all SMs, we use a large number of blocks to access the L2 cache. We then calculate the bandwidth of the L2 cache based on the amount of data accessed and the time consumed.
\subsubsection{Global memory} 
For the latency test, we first allocate a global memory that exceeds the L2 size to avoid L2 pre-fetching, and then initialize the global memory. Initialization has two purposes. The first is to enable the test to be performed at a fixed stride, and the second is to warm up the TLB to avoid the occurrence of cold misses. When we started the test, we launched four consecutive threads, each of which was responsible for reading 8 bytes, thus forming a 32-byte memory read transaction. Finally, we can calculate the memory access latency of each thread.
For the throughput test, we allocate a much larger memory space than L2. We set up each thread to use vectorized memory access to read float4. Each thread reads 5 times and writes 1 time. Finally, we calculate the memory bandwidth based on the time consumed and the amount of data.

\subsection{Tensor Core Latency and Throughput}
\subsubsection{Tensor Core's Evolution}\label{sec:evo_tc}
Table \ref{tab:TCS} illustrates the progression of TCs, encompassing enhancements in precision, operand shapes, programming modes, and execution modes. In the initial Volta Architecture, first-generation TCs exclusively supported FP16 as the input data type. Subsequent architectures, including Ampere, Ada, and Hopper, introduced support for a broader range of data types such as BF16, TF32, FP64, INT8, INT4, Binary, and more.
The programming of TCs has also seen continuous improvement. Ampere and Ada Lovelace GPUs provide users with the flexibility to utilize either the legacy C-level \emph{wmma} APIs or PTX-level \emph{mma} instructions. Notably, the \emph{wmma} APIs had limitations in fully harnessing TCs' capabilities, whereas \emph{mma} instructions could leverage advanced sparse matrix multiplication capabilities introduced since Ampere. In the case of Hopper GPUs, new warp-group-level \emph{wgmma} instructions were introduced. Both \emph{wmma} and \emph{mma} APIs remain supported in Hopper, but we find that the complete potential of Hopper TCs can only be realized through \emph{wgmma} instructions.

\begin{figure*}[htbp]
    \centering
    \includegraphics[width=0.9\linewidth]{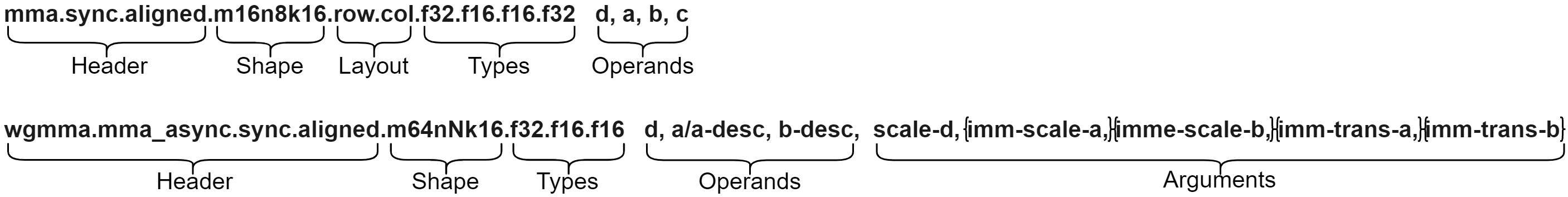}  
    \caption{The \emph{mma} and \emph{wgmma} instructions that perform $D = A \times B + C$ and $D = A \times B \{+ D\}$, respectively.}
    \label{fig:wgmma_inst}
\end{figure*}

Fig. \ref{fig:wgmma_inst} provides examples of both \emph{mma} and \emph{wgmma} instructions, demonstrating mixed-precision capabilities. An \emph{mma} instruction computes $D(m \times n) = A(m \times k) \times B(k \times n) + C(m \times n)$ and is executed by one CUDA warp (i.e., 32 threads) synchronously. In contrast, \emph{wgmma} for Hopper computes $D(m \times n) = A(m \times k) \times B(k \times n) + D(m \times n)$ and is asynchronously executed by one CUDA warp group (i.e., four CUDA warps). The matrix shapes for \emph{mma} instructions can be $m16n8k16$ or $m16n8k8$, while \emph{wgmma} supports $m64nNk16$, where $N$ can be 16, 32, 64, 128, 256, and so on, with the complete valid range listed in \cite{nvidia-ptx-doc}.
Notably, \emph{wgmma} has the advantage of directly loading matrices $A$ and $B$ from shared memory, unlike \emph{mma}, which requires storing all matrices in the register file before execution. We use the term ``SS" to denote the \emph{wgmma} instruction that loads both $A$ and $B$ from shared memory, while ``RS" is used for the instruction that loads $A$ from the register file. Additionally, \emph{wgmma} offers support for certain useful arguments not required for \emph{mma}. Further details are provided in \cite{nvidia-ptx-doc}.


\begin{table}[!ht]
\caption{Properties of the latest generations of Tensor Cores}
\label{tab:TCS}
\begin{tabular}{|l|l|l|l|}
\hline
\textbf{Arch} & \textbf{Precision} & \textbf{Programmability} & \textbf{Mode} \\ \hline
Ampere & \begin{tabular}[c]{@{}l@{}}FP16,BF16,\\ TF32,FP64,\\ INT8,INT4,Binary\end{tabular} & \begin{tabular}[c]{@{}l@{}}C: wmma\\ PTX: mma, mma.sp\end{tabular} & Sync \\ \hline
Ada & \begin{tabular}[c]{@{}l@{}}FP16,BF16,FP8,\\ TF32,FP64,\\ INT8,INT4,Binary\end{tabular} & \begin{tabular}[c]{@{}l@{}}C: wmma\\ PTX:mma, mma.sp\end{tabular} & Sync \\ \hline
\multirow{2}{*}{Hopper} & \multirow{2}{*}{\begin{tabular}[c]{@{}l@{}}FP16,BF16,FP8,TF32,\\ FP64,INT8,Binary\end{tabular}} & \begin{tabular}[c]{@{}l@{}}C: wmma \\ PTX: mma, mma.sp\end{tabular} & Sync \\ \cline{3-4} 
 &  & PTX: wgmma, wgmma.sp & ASync \\ \hline
\end{tabular}
\end{table}

\subsubsection{Benchmarking Levels and Performance Metrics}
We conduct micro-benchmarking of TCs at the PTX level, as it strikes a suitable balance between granularity and complexity. Additionally, we disassemble PTX instructions to SASS codes to achieve a deeper understanding of the operations.
We focus on assessing two critical indicators: latency and throughput. Latency signifies the elapsed time, measured in clock cycles, starting from the initiation of instruction issuance to the execution pipeline and concluding when the results become accessible for subsequent usage. This measurement is specifically labeled as ``completion latency." To elaborate further, we issue a single synchronous TC instruction (i.e., \emph{mma}) using one CUDA warp per SM, whereas one asynchronous TC instruction (i.e., \emph{wgmma}) is issued utilizing four CUDA warps (comprising a warp group) on one SM. We execute the instruction 1024 times within a CUDA kernel. Throughput is quantified as $Total\_{OPS}/Duration$, where $OPS$ represents multiplication or addition operations. It's important to emphasize that, unlike the approach described in \cite{sun_tpds_2023}, we abstain from utilizing total clock cycles to compute throughput due to potential variations in GPU frequencies during the execution of different TC instructions.


\subsection{Transformer Engine}
The Transformer Engine (TE) \cite{te} is a library specifically designed to accelerate Transformer models\cite{vaswani2017attention}, following the introduction of the Hopper architecture. It is capable of leveraging the FP8 precision offered by both the Hopper and Ada architectures. In particular, it provides a variety of optimized modules for Transformer layers that can be utilized within the widely-used deep learning framework, PyTorch \cite{pytorch}. In this subsection, we describe the benchmark details of the Transformer Engine on different modules, as well as the inference performance of FP8 in Large Language Models.

\subsubsection{Linear Layer}
In the Transformer architecture, most of the computational overhead comes from the linear layers, specifically matrix multiplication while the Transformer Engine provides the \texttt{te.Linear} implementation to perform matrix multiplication with higher throughput on FP8 Tensor Cores.
When employing the Transformer Engine with \texttt{te.Linear} for matrix multiplication in FP8 precision, TE converts both the input and weights in the linear layer to FP8. This conversion process involves data transformation and quantization operations. For example, as the dynamic range of FP8 may not encompass the maximum value of the input tensor, TE identifies the maximum absolute value of the input data as the scaling factor. It then adjusts the input data to fit the representation range of FP8 using $inp_{fp8} = inp_{fp16} / scale$, followed by matrix multiplication in FP8 Tensor Core $out_{fp8} = inp_{fp8} \times w_{fp8}$. Finally, it scales the result with $out_{fp16} = out_{fp8} \times scale$. This operation would introduce some overhead.

\begin{figure}[htbp]
    \centering
    \includegraphics[width=0.9\linewidth]{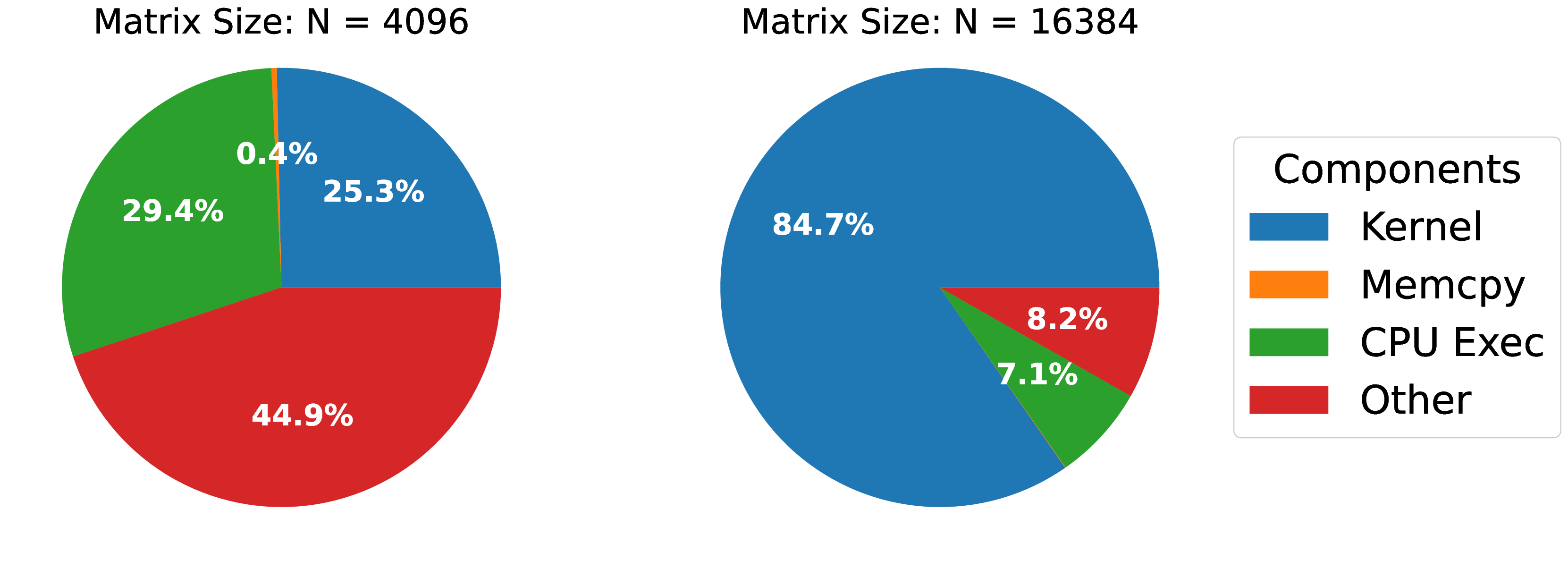} 
    \caption{Proportion of execution time for different operators when performing FP8 matrix multiplication using \texttt{te.Linear}.}
    \label{fig:Linear_pie}
\end{figure}

As depicted in Fig. \ref{fig:Linear_pie}, when performing matrix multiplication with relatively small matrix sizes, the proportion of overhead attributed to conversion is significantly larger than that resulting from the GEMM kernel computation in FP8 Tensor Cores.
To evaluate and investigate the support and optimization of TE for linear layers, we measure the throughput (GFLOPS) of \texttt{te.Linear} for two identical matrices $D(n \times n) = A(n \times n) \times B(n \times n)$. 

\subsubsection{TransformerLayer}
The Transformer Engine (TE) capitalizes on the efficiency improvements provided by FP8 through specific operator fusion optimizations for transformer layer structures. For example, \texttt{te.LayerNormMLP} combines layernorm and MLP within the transformer structure, allowing data transmission between layernorm and the subsequent MLP layer to adopt the FP8 format. This approach not only eliminates data format conversion overhead but also effectively leverages FP8 memory transfer advantages.

TE offers a \texttt{te.TransformerLayer} module that encompasses all operator optimizations for transformer layer structures, facilitating the implementation of various Large Language Model (LLM) structures by adjusting its parameters. However, some operators, such as \texttt{Softmax} and \texttt{GeLU}, have not been quantized to FP8 by TE, resulting in significant data format conversion overhead. Additionally, the DotProductAttention operator uses flash-attention \cite{flashattention} rather than FP8 Tensor Cores.

The computational overhead of the transformer layer's linear layer primarily depends on the hidden size, raising the question of which hidden state (dimension of embedding) will yield better performance for TE with FP8 compared to FP16. We investigate this by examining the open-source LLM, Llama \cite{llama1, llama2}, modifying the activation function to SwiGLU \cite{swiglu} and normalization to RMSNorm \cite{rmsnorm}. We set layer structure parameters based on the hidden state's size, with hidden states 4096, 5120, and 8192 corresponding to Llama configurations 7b, 13b, and 70b, respectively.

\begin{table}[!ht]
\caption{Parameter settings of \texttt{te.TransformerLayer} for various $hidden\_size$ values}
    \centering
    \begin{tabular}{|l|l|l|l|l|l|}
    \hline
        hidden\_size & 1024 & 2048 & 4096 & 5120 & 8192 \\ \hline
        ffn\_hidden\_size & 2816 & 5632 & 11008 & 13824 & 22016 \\ \hline
        num\_attention\_heads & 8 & 16 & 32 & 40 & 64 \\ \hline
    \end{tabular}
    \label{tab:telayer}
\end{table}

We fixed the input as (4, 512, $hidden\_size$), where 4 is the $batch\_size$, 512 is the $sequence\_length$, and the attention mask is set to None. We then calculated the latency (ms) required for encoding a single layer once, focusing on the encoding task for a single layer. 

\subsubsection{LLM Generation}
Currently, the Transformer Engine has not provided optimal support for mainstream decode-only casual language models. In order to test the inference performance of the Transformer Engine on this type of model like, Llama \cite{llama1}, we replaced the \texttt{nn.Linear} and \texttt{RMSNorm} in the original model structure with \texttt{te.Linear} and \texttt{te.RMSNorm}, respectively, to ensure that most modules in the model utilize the Transformer Engine.

To evaluate the effectiveness of TE in generating text for Llama, we followed \cite{vllm} using the ShareGPT dataset as input for the LLM. The ShareGPT dataset comprises conversations between users and ChatGPT \cite{chatgpt}, which have been shared by the users. We tokenize these datasets and, based on their input and output lengths, generate synthesized client requests.

In order to test and ensure compatibility with different hardware architectures (different memory capacities), we set the maximum input length to 128 and the maximum text generation length to 128. Furthermore, to meet the dimension requirements of \texttt{Te.Linear}, we set the batch size to 8.

We use throughput as the evaluation metric, which represents the total text length that can be processed per second:
$Throughput = (input\_len + output\_len) / time$.

\subsection{New CUDA Programming Features}
\subsubsection{DPX}
Nvidia offers DPX functions\footnote{\url{https://docs.nvidia.com/cuda/cuda-c-programming-guide/index.html#dpx}} from CUDA 12 onward to accelerate dynamic programming code, enhancing programming ease. On the latest Hopper architecture, these functions are hardware-accelerated. Our test of DPX functions focuses on the instruction latency and throughput.
For latency assessment, we utilize a thread to iteratively issue DPX functions, calculating their average latency. In the throughput test, we employ a block to repeatedly issue DPX functions, determining the DPX instruction throughput for each SM.
To pinpoint the location of DPX acceleration hardware, we vary the number of launched blocks and observed the relationship between DPX throughput and the launched block count.
\subsubsection{Asynchronous Data Movement}
The introduction of asynchronous execution is a highlight of the Ampere architecture. This feature allows for non-blocking data transfers between GPU global memory and shared memory using \emph{cuda::memcpy\_async}, avoiding thread occupation during data movement. It facilitates the overlap of computations with data transfers, effectively reducing overall execution time. Building upon Ampere's asynchronous copies, the Hopper architecture enhances this with a more advanced Tensor Memory Accelerator (TMA) for sophisticated asynchronous copying.

To assess this feature's efficiency, we conduct empirical studies using the \textsl{globalToShmemAsyncCopy} application from official CUDA samples\footnote{\url{https://github.com/NVIDIA/cuda-samples/Samples/3_CUDA_Features/globalToShmemAsyncCopy}}. The application implements matrix multiplication and leverages asynchronous data copy from global to shared memory for compute capability 8.0 or higher.
We compare two implementations: ``SyncShare'', employing synchronous copy to shared memory for conventional tiled matrix multiplication, and ``AsyncPipe'', enhancing tiling with asynchronous data movement. The asynchronous version uses a two-stage pipeline with doubled shared memory buffer size, enabling computation and data copy overlap across different execution streams.
Matrix A's width and Matrix B's height are set to 2048, determining each thread's computational workload. We vary the block size from 8$\times$8 to 32$\times$32 to assess asynchronous operations' effects on warp concurrency. Additionally, we benchmark different block numbers to optimize computational throughput by adjusting Matrix A's height and Matrix B's width.
\subsubsection{Distributed Shared Memory}
The Hopper architecture features a direct SM-to-SM communication network within clusters, enabling threads from one thread block to access shared memory of another block, known as distributed shared memory (DSM). According to the official documentation, this network can reduce data transfer overhead between blocks on different SMs by up to 7$\times$. Additionally, for cases where shared memory demand restricts active block numbers on an SM, DSM can partition data within the same cluster, alleviating shared memory demand per block.
The programmability of DSM is facilitated through the CUDA C function, \texttt{cluster.map\_shared\_rank(SMEM, DST\_BLOCK\_RANK)}, returning the shared memory address of the target block. Here, \texttt{SMEM} represents the shared memory pointer, and \texttt{DST\_BLOCK\_RANK} is the target block rank in the cluster. This is compiled into PTX code \emph{mapa}, which maps the address of the shared variable in the target block.

We assess DSM using three benchmarks. 

(1) Latency Measurement:
To measure inter-SM data transfer latency, we launch two blocks, each with one thread. Using the \emph{clock()} function, we record the latency of adding register values of the first block to the second one. We utilize \emph{mov.u32 \%0, \%\%smid} PTX code to record the SM ID of each block, ensuring they run on different SMs.

(2) Throughput Measurement by RBC (Ring-Based Copy):
For DSM data communication throughput, we propose the ring-based copy (RBC) scheme. We launch one block per SM and gather them into clusters. In each cluster, we arrange threads in each block ranked by $R$ to add their register values to those of the block ranked by $(R+1)\%CS$, where $CS$ refers to the cluster size. Instruction-level parallelism (ILP) is employed to maximize bandwidth. We tune cluster size, block size, and ILP to measure their impact on DSM throughput.

(3) Histogram Application with DSM:
We redesign the histogram\footnote{\url{https://github.com/NVIDIA/cuda-samples/Samples/2_Concepts_and_Techniques/histogram}} application using DSM, distributing bins across blocks in the same cluster. During histogram counting via shared memory, each thread loads the element and determines the DSM address for the target bin, followed by an atomic increment operation. We adjust cluster size, block size, and bin count, measuring element processing throughput.

\section{Experimental Results}
In this section, we first introduce the GPUs used in this study. Then, we introduce and analyze the performance of memory, Tensor Core, AI, and new CUDA features. Due to space limitations, we hereby present the most meaningful findings. The complete experimental results can be found in the public preprint version or reproduced by the open-sourced codes after the review process.
\subsection{Experimental Setup}

In this work, we select the most representative GPUs of the Ampere, Ada Lovelace, and Hopper architectures, which are A100 PCIe, RTX4090, and H800 PCIe respectively. Their basic hardware properties are shown in Table \ref{tab:hw_device}. On the RTX4090, the driver version we use is 530.30.02 and the CUDA version is 12.1. On A100 and H800, the driver version we use is 535.104.05, and the CUDA version is 12.2. 

\begin{table}[]
\caption{Comparison of the Properties of the Ampere, Ada Lovelace and Hopper Devices}
\label{tab:hw_device}
\centering
\begin{tabular}{|l|c|c|c|}
\hline
\textbf{Device}                                                      & \textbf{A100 PCIe}                                        & \textbf{RTX4090}                                              & \textbf{H800 PCIe}                                        \\ \hline
Comp. Capability                                                     & \begin{tabular}[c]{@{}c@{}}8.0 \\ (Ampere)\end{tabular}   & \begin{tabular}[c]{@{}c@{}}8.9 \\ (Ada Lovelace)\end{tabular} & \begin{tabular}[c]{@{}c@{}}9.0 \\ (Hopper)\end{tabular}   \\ \hline
SMs * cores/SM                                                       & 108 * 64                                                  & 128 * 128                                                     & 114 * 128                                                 \\ \hline
Max Clock rate                                                       & 1410 MHz                                                  & 2520 MHz                                                      & 1755 MHz                                                  \\ \hline
Mem. Size                                                            & 40GB                                                      & 24GB                                                          & 80GB                                                      \\ \hline
Mem. Type                                                            & HBM2e                                                     & GDDR6X                                                        & HBM2e                                                     \\ \hline
Mem. Clock rate                                                      & 1215 Mhz                                                  & 10501 Mhz                                                     & 1593 Mhz                                                  \\ \hline
Mem. Bus                                                             & 5120-bit                                                  & 384-bit                                                       & 5120-bit                                                  \\ \hline
Mem. Bandwidth                                                       & 1555 GB/s                                                 & 1008 GB/s                                                     & 2039 GB/s                                                 \\ \hline
Tensor Core                                                          & \begin{tabular}[c]{@{}c@{}}432 \\ (3rd Gen.)\end{tabular} & \begin{tabular}[c]{@{}c@{}}512 \\ (4th Gen.)\end{tabular}     & \begin{tabular}[c]{@{}c@{}}456 \\ (4th Gen.)\end{tabular} \\ \hline
DPX hardware                                                         & \multirow{3}{*}{No}                                       & \multirow{3}{*}{No}                                           & \multirow{3}{*}{Yes}                                      \\ \cline{1-1}

\begin{tabular}[c]{@{}l@{}}Distributed \\ shared memory\end{tabular} &                                                           &                                                               &                                                           \\ \hline

\end{tabular}
\end{table}

\subsection{Memory Latencies and Throughputs}

Table \ref{tab:mem_lat} shows the memory access latency at different memory levels. We can observe that on devices with different architectures, their memory access latencies are close, indicating that the memory levels are similar. But we can find that the global memory latency of A100 and H800 using HBM2e will be slightly lower than RTX4090. In the comparison of latency at different memory levels, it can be observed that on the three devices, the average latency of the L2 cache is 6.5 times that of the L1 cache, and the average latency of the global memory is 1.9 times that of the L2 cache.

\begin{table}[]
\caption{Latency clocks of different memory scopes}
\label{tab:mem_lat}
\centering
\begin{tabular}{|c|c|c|c|}
\hline
\textbf{Type} & \textbf{RTX4090} & \textbf{A100} & \textbf{H800} \\ \hline
L1 Cache      & 43.4             & 37.9          & 40.7          \\ \hline
Shared & 30.1             & 29.0          & 29.0          \\ \hline
L2 Cache      & 273.0            & 261.5         & 263.0         \\ \hline
Global & 541.5            & 466.3         & 478.8         \\ \hline
\end{tabular}
\end{table}

Table \ref{tab:mem_bw} shows the memory access throughput at different memory levels. In the cache throughput test, we use different data types for memory access. We can observe that using vectorized memory access (FP32.v4, equivalent to CUDA's float4) can always achieve better performance. It is worth noting that in FP64 Cache access, the throughput of RTX4090 and H800 will be much smaller than the normal value. This is because we need to perform calculations after accessing memory to avoid the elimination of our memory access instructions by the compiler. However, the FP64 addition throughput of both RTX4090 and H800 we measured is 16byte/clk/SM. Therefore, the bottleneck of the memory access test here lies in the FP64 computing unit, which does not represent the actual throughput of the cache memory.

The maximum throughput of L1 Cache and shared memory of the three devices are similar. However, for the throughput of L2 Cache, H800 is 2.6 times and 2.2 times that of RTX4090 and A100 respectively.\footnote{Note that for the L1 cache, the amount of data they transfer per clock is almost the same. However, since the order of clock frequency from high to low is RTX4090, H800, and A100, the order of throughput per unit time from high to low is also RTX4090, H800, and A100. The same calculation method also applies to L2 cache.} In the memory throughput test, our results reach 92\%, 90\%, and 91\% of the theoretical performance on RTX4090, A100, and H800 respectively. In the comparison of L2 and Global, the L2 cache throughput of RTX4090, A100, and H800 is 4.67, 2.01, and 4.23 times the global memory throughput respectively.
\begin{table*}[]
\caption{Throughput at different memory levels}
\label{tab:mem_bw}
\centering
\begin{tabular}{|c|ccc|ccc|ccc|}
\hline
\textbf{Type}                                                                      & \multicolumn{3}{c|}{\textbf{RTX4090}}                                                      & \multicolumn{3}{c|}{\textbf{A100}}                                                         & \multicolumn{3}{c|}{\textbf{H800}}                                                         \\ \hline
\multirow{2}{*}{\begin{tabular}[c]{@{}c@{}}L1 Cache (byte/clk/SM)\end{tabular}} & \multicolumn{1}{c|}{FP32} & \multicolumn{1}{c|}{FP64} & FP32.v4 & \multicolumn{1}{c|}{FP32} & \multicolumn{1}{c|}{FP64} & FP32.v4 & \multicolumn{1}{c|}{FP32} & \multicolumn{1}{c|}{FP64} & FP32.v4 \\ \cline{2-10} 
                                                                                   & \multicolumn{1}{c|}{63.7}          & \multicolumn{1}{c|}{13.3}          & 121.2            & \multicolumn{1}{c|}{99.5}          & \multicolumn{1}{c|}{120.0}         & 106.8            & \multicolumn{1}{c|}{125.8}         & \multicolumn{1}{c|}{16.0}          & 124.1            \\ \hline

\multirow{2}{*}{\begin{tabular}[c]{@{}c@{}}L2 Cache (byte/clk)\end{tabular}}  & \multicolumn{1}{c|}{FP32} & \multicolumn{1}{c|}{FP64} & FP32.v4 & \multicolumn{1}{c|}{FP32} & \multicolumn{1}{c|}{FP64} & FP32.v4 & \multicolumn{1}{c|}{FP32} & \multicolumn{1}{c|}{FP64} & FP32.v4 \\ \cline{2-10} 
                                                                                   & \multicolumn{1}{c|}{1622.2}        & \multicolumn{1}{c|}{1500.8}        & 1708.0           & \multicolumn{1}{c|}{1853.7}        & \multicolumn{1}{c|}{1990.4}        & 2007.9           & \multicolumn{1}{c|}{4472.3}        & \multicolumn{1}{c|}{1817.3}        & 3942.4           \\ \hline
\begin{tabular}[c]{@{}c@{}}Shared Memory (byte/clk/SM)\end{tabular}                     & \multicolumn{3}{c|}{127.9}                                                                 & \multicolumn{3}{c|}{128.0}                                                                 & \multicolumn{3}{c|}{127.9}                                                                 \\ \hline
\begin{tabular}[c]{@{}c@{}}Global Memory (GB/s)\end{tabular}                                                                     & \multicolumn{3}{c|}{929.8}                                                                 & \multicolumn{3}{c|}{1407.2}                                                                & \multicolumn{3}{c|}{1861.5}                                                                \\ \hline
\begin{tabular}[c]{@{}c@{}}L2 vs. Global \end{tabular}                                                                     & \multicolumn{3}{c|}{4.67×}                                                                 & \multicolumn{3}{c|}{2.01×}                                                                & \multicolumn{3}{c|}{4.23×}                                                                \\ \hline
\end{tabular}
\end{table*}

\subsection{Tensor Core Latencies and Throughputs}
\noindent\textbf{SASS analysis.} We perform the disassembly of \emph{mma} and \emph{wgmma} instructions specifically for Hopper Tensor Cores, and the results are presented in Table \ref{tab:sassanalysis}. The \emph{mma} instructions undergo compilation into SASS instructions, with the naming convention following the established patterns: HMMA (for floating-point types), IMMA (for integer types), and BMMA (for binary types). It is worth noting the existence of two specialized types within \emph{mma}: INT4 and FP8.

For INT4, on Ampere and Ada Tensor Cores, \emph{mma} instructions are compiled into IMMA.16832.S4.S4 instructions. However, a noteworthy deviation occurs on Hopper, where INT4 \emph{mma} instructions are compiled into a series of IMAD instructions, which eventually run on the CUDA cores. This deviation results in performance that may fall short of the expected performance levels achievable with Tensor Cores. Additionally, there are no \emph{mma} instructions available for FP8, a new data type introduced in Ada.

The latest \emph{wgmma} instructions are currently exclusive to Hopper Tensor Cores, despite Nvidia's assertion that both Ada and Hopper feature 4th generation Tensor Cores. Unlike \emph{mma}, \emph{wgmma} instructions are compiled into the new GMMA SASS instructions. Users have the capability to program two variations of FP8 (E5M2 and E4M3) Tensor Cores using \emph{wgmma}. It's important to note that \emph{wgmma} does not offer support for INT4 Tensor Cores.
\begin{table}[]
\addtolength{\tabcolsep}{-3.5pt}
\caption{SASS Instructions for Different Hopper Tensor Core PTX Instructions}
\label{tab:sassanalysis}
\centering
\scriptsize
\begin{tabular}{|l|l|l|l|}
\hline
\textbf{A/B} & \textbf{C/D} & \textbf{mma} & \textbf{wgmma} \\ \hline
FP16 & FP16 & HMMA.16816.F16 & HGMMA.64x256x16.F16 \\ \hline
FP16 & FP32 & HMMA.16816.F32 & HGMMA.64x256x16.F32 \\ \hline
TF32 & FP32 & HMMA.1688.F32.TF32 & HGMMA.64x256x8.F32.TF32 \\ \hline
FP8 & FP16 & $\times$ & \begin{tabular}[c]{@{}l@{}}QGMMA.64x256x32.F16.E5M2.E5M2\\ QGMMA.64x256x32.F16.E4M3.E4M3\end{tabular} \\ \hline
FP8 & FP32 & $\times$ & \begin{tabular}[c]{@{}l@{}}QGMMA.64x256x32.F32.E4M3.E4M3\\ QGMMA.64x256x32.F32.E5M2.E5M2\end{tabular} \\ \hline
INT8 & INT32 & IMMA.16832.S8.S8 & IGMMA.64x256x32.S8.S8 \\ \hline
INT4 & INT32 & \begin{tabular}[c]{@{}l@{}}IMAD.MOV.U32 \end{tabular} & $\times$ \\ \hline
Binary & INT32 & BMMA.168256.AND.POPC & BGMMA.64x256x256.AND.POPC \\ \hline
\end{tabular}
\end{table}
    
\noindent\textbf{\emph{mma} results.} Table \ref{tab:syncwgmma} provides an overview of the latency and throughput measurements for \emph{mma} instructions across A100, RTX4090, and H800 Tensor Cores GPUs. Note that the sparse shapes in the table represent compressed shapes. In other words, the $k$ of the actual instruction modifier is twice that in the table.

For A100 and H800, the same-precision \emph{mma} instructions with the larger shapes commonly achieve better throughputs. But this phenomenon disappears on RTX4090. Sparse and dense \emph{mma} instructions exhibit equivalent latency, with sparse \emph{mma} instructions achieving higher throughputs. On the RTX4090, sparse \emph{mma} instructions can achieve up to double the throughput compared to their corresponding dense counterparts, aligning with the speedup claims stated in the vendor's documentation. However, for the A100, only the sparse \emph{mma} instructions with larger shapes can realize the theoretical speedups. In the case of the H800, sparse \emph{mma} instructions can only achieve an average speedup of 1.42 times over the dense ones. This highlights that on Hopper Tensor Cores, sparse \emph{mma} instructions may not fully harness the capabilities of the sparse tensor cores.

The achieved throughput on A100 exceed 95\% of their theoretical peak performance. The achieved throughput of RTX4090 is higher than the official theoretical peak performance. This is because we find that our RTX4090 runs at a higher frequency than the officially announced boost frequency. However, on Hopper Tensor Cores, \emph{mma} instructions can only attain an average of 62.9\% of the theoretical peak performance. When developing high-performance applications tailored for Hopper GPUs, such as matrix multiplication and convolution, users should exercise caution in their utilization of \emph{mma} instructions.

\begin{table*}[]
\caption{Different dense and sparse \emph{mma} instructions on A100, RTX4090 and H800 Tensor Cores. Latency (LAT) is measured in clock cycles. Throughput is measured in TFLOPS or TOPS/s. Peak performance (A100): FP16 (312 TFLOPS); TF32 (156 TFLOPS); INT8 (624 TOPS). Peak performance (RTX4090): FP16 (330.3 TFLOPS); TF32 (82.6 TFLOPS); INT8 (660.6 TOPS).  Peak performance (H800): FP16 (756.5 TFLOPS); TF32 (378 TFLOPS); INT8 (1513 TOPS).}
\label{tab:syncwgmma}
\begin{tabularx}{\textwidth}{|X|X|X|XXXXXX|}
\hline
\multirow{3}{*}{\textbf{A/B}} & \multirow{3}{*}{\textbf{C/D}} & \multirow{3}{*}{\textbf{Shape}} & \multicolumn{6}{c|}{\textbf{LAT/Throughput}} \\ \cline{4-9} 
 &  &  & \multicolumn{2}{c|}{\textbf{A100}} & \multicolumn{2}{c|}{\textbf{RTX4090}} & \multicolumn{2}{c|}{\textbf{H800}} \\ \cline{4-9} 
 &  &  & \multicolumn{1}{l|}{Dense} & \multicolumn{1}{l|}{Sparse} & \multicolumn{1}{l|}{Dense} & \multicolumn{1}{l|}{Sparse} & \multicolumn{1}{l|}{Dense} & Sparse \\ \hline
FP16 & FP16 & m16n8k8 & \multicolumn{1}{l|}{17.7/310.0} & \multicolumn{1}{l|}{17.3/408.4} & \multicolumn{1}{l|}{17.7/355.3} & \multicolumn{1}{l|}{17.3/713.2} & \multicolumn{1}{l|}{16.0/368.6} & 16.0/493.8 \\ \hline
FP16 & FP16 & m16n8k16 & \multicolumn{1}{l|}{24.6/310.6} & \multicolumn{1}{l|}{24.5/622.8} & \multicolumn{1}{l|}{24.6/357.6} & \multicolumn{1}{l|}{24.5/711.8} & \multicolumn{1}{l|}{24.1/494.4} & 24.0/722.8 \\ \hline
FP16 & FP32 & m16n8k8 & \multicolumn{1}{l|}{17.5/299.6} & \multicolumn{1}{l|}{18.0/394.1} & \multicolumn{1}{l|}{18.8/177.8} & \multicolumn{1}{l|}{18.8/357.4} & \multicolumn{1}{l|}{16.0/363.7} & 16.0/488.7 \\ \hline
FP16 & FP32 & m16n8k16 & \multicolumn{1}{l|}{26.0/303.4} & \multicolumn{1}{l|}{24.5/603.3} & \multicolumn{1}{l|}{33.0/178.9} & \multicolumn{1}{l|}{33.0/356.0}
& \multicolumn{1}{l|}{24.1/490.7} & 24.0/721.8 \\ \hline
TF32 & FP32 & m16n8k4 & \multicolumn{1}{l|}{17.8/149.5} & \multicolumn{1}{l|}{18.2/196.8} & \multicolumn{1}{l|}{19.2/89.0} & \multicolumn{1}{l|}{19.0/178.0} & \multicolumn{1}{l|}{16.5/180.6} & 16.4/240.7 \\ \hline
TF32 & FP32 & m16n8k8 & \multicolumn{1}{l|}{26.3/151.5} & \multicolumn{1}{l|}{26.7/301.5} & \multicolumn{1}{l|}{33.4/89.0} & \multicolumn{1}{l|}{33.3/178.7} & \multicolumn{1}{l|}{24.5/246.4} & 24.4/363.3 \\ \hline
INT8 & INT32 & m16n8k16 & \multicolumn{1}{l|}{17.6/594.8} & \multicolumn{1}{l|}{18.0/788.5} & \multicolumn{1}{l|}{17.3/707.6} & \multicolumn{1}{l|}{17.3/1412} & \multicolumn{1}{l|}{16.1/730.3} & 16.1/970.0 \\ \hline
INT8 & INT32 & m16n8k32 & \multicolumn{1}{l|}{26.0/607.6} & \multicolumn{1}{l|}{26.6/1210} & \multicolumn{1}{l|}{24.5/711.7} & \multicolumn{1}{l|}{24.6/1423} & \multicolumn{1}{l|}{24.0/977.9} & 24.2/1435 \\ \hline
  \end{tabularx}
\end{table*}
   
\noindent\textbf{\emph{wgmma} results.} As a set of warp-group-level Tensor Core instructions designed specifically for Hopper GPUs, \emph{wgmma} instructions are the pioneering instructions to be executed asynchronously. Table \ref{tab:densewgmma} and \ref{tab:sparsewgmma} show the measured latency and throughputs of dense and sparse \emph{wgmma} instructions, respectively. When initializing matrices with zeros, we achieve throughputs exceeding 95\% of the theoretical peak performance. We observe a decrease in Tensor Core performance when initializing matrices with random values, especially pronounced when utilizing FP16 as the computation type and FP32 for accumulation. This phenomenon is primarily attributed to the power consumption nearing the 350W power limit of the H800-PCIe, subsequently causing a reduction in frequency. Users working with Tensor Cores on the H800-PCIe GPU should take into full consideration the power constraints when performing computations. 

In the case of dense \emph{wgmma}, with $N$ set to 128, we observe that the latency for all data types corresponding to the instructions is 128.0. Interestingly, under both ``RS'' and ``SS'' modes, the latency and throughputs for the same instruction remain relatively consistent. We attribute this phenomenon to the efficient concealment of shared memory access latency due to the substantial computational workload and asynchronous nature of the process.

In the context of sparse \emph{wgmma} instructions, the latencies for ``RS'' and ``SS'' modes are 128.0 and 144.0, respectively. Additionally, we observe that the achieved throughputs in ``SS'' modes are lower compared to ``RS'' modes, which is notably different from dense \emph{wgmma}. We find that in sparse \emph{wgmma}, the ``SS'' mode retrieves data from the shared memory of size $m \times k$ and performs a 2:4 sparse pruning based on metadata during the execution of the sparse \emph{wgmma} instruction. In contrast, the ``RS'' mode directly accesses data from the pruned register file of size $m \times k / 2$. The high shared memory access demand (twice as much) may lead to latency that cannot be effectively concealed by Tensor Core computation, resulting in sparse \emph{wgmma} instructions in ``SS'' modes failing to achieve the expected peak performance.

  \begin{table*}[!ht]
   \caption{Variations in Dense \emph{wgmma} Instructions for H800 Tensor Cores. Latency (LAT) is quantified in clock cycles, while throughput is expressed in TFLOPS or TOPS. The peak throughputs for FP16, TF32, FP8, and INT8 are 756.5, 373, 1513, and 1513, correspondingly. ``Zero" or ``Rand" signifies that all matrices are initialized with either zero or randomly generated values. ``SS" implies that both matrix A and B are stored in shared memory, while ``RS'' signifies that matrix A is stored in the register file, whereas B is stored in shared memory. ``Rand" has the same latency as ``Zero".}
   \label{tab:densewgmma}
   \begin{tabularx}{\textwidth}{|X|X|X|X|X|X|X|}
   \hline
   \textbf{A/B} &
   \textbf{C/D} &
   \textbf{Instruction} &
   \textbf{\begin{tabular}[c]{@{}l@{}}LAT/Throughput\\ (SS,Zero)\end{tabular}} &
   \textbf{\begin{tabular}[c]{@{}l@{}}LAT/Throughput\\ (RS,Zero)\end{tabular}} &
   \textbf{\begin{tabular}[c]{@{}l@{}}Throughput\\ (SS,Rand)\end{tabular}} &
   \textbf{\begin{tabular}[c]{@{}l@{}}Throughput\\ (RS,Rand)\end{tabular}} \\ \hline
   FP16 & FP16  & m64n256k16 & 128.0/729.3  & 128.0/729.2  & 704.5  & 703.7  \\ \hline
   FP16 & FP32  & m64n256k16 & 128.0/728.5  & 128.0/731.9  & 665.4  & 667.5  \\ \hline
   TF32 & FP32  & m64n256k8  & 128.0/364.4  & 128.0/364.6  & 357.1  & 357.3  \\ \hline
   FP8  & FP16  & m64n256k32 & 128.0/1448.4 & 128.0/1448.0 & 1439.2 & 1440.3 \\ \hline
   FP8  & FP32  & m64n256k32 & 128.0/1447.5 & 128.0/1455.0 & 1417.2 & 1419.8 \\ \hline
   INT8 & INT32 & m64n256k32 & 128.0/1448.7 & 128.0/1447.9 & 1442.3 & 1442.2 \\ \hline
   \end{tabularx}
   \end{table*}

    \begin{table*}[]
    \caption{Different sparse \emph{wgmma} Instructions on H800 Tensor Cores. The definitions of LAT and throughput can be found in the caption of Table \ref{tab:densewgmma}. ``Rand" has the same latency as ``Zero".}
    \label{tab:sparsewgmma}
     \begin{tabularx}{\textwidth}{|X|X|X|X|X|X|X|}
     \hline
      \textbf{A/B} & \textbf{C/D} & \textbf{\begin{tabular}[c]{@{}l@{}}Sparse \\ Instruction\end{tabular}} & \textbf{\begin{tabular}[c]{@{}l@{}}LAT/Throughput\\ (SS,Zero)\end{tabular}} & \textbf{\begin{tabular}[c]{@{}l@{}}LAT/Throughput\\ (RS,Zero)\end{tabular}} & \textbf{\begin{tabular}[c]{@{}l@{}}Throughput\\ (SS,Rand)\end{tabular}} & \textbf{\begin{tabular}[c]{@{}l@{}}Throughput\\ (RS,Rand)\end{tabular}} \\ \hline
       FP16 & FP16 & sp.m64n256k32 & 144.0/1308.0 & 128.0/1472.0 & 1257.8 & 1362.3 \\ \hline
       FP16 & FP32 & sp.m64n256k32 & 144.0/1312.3 & 128.0/1476.2 & 1194.3 & 1277.5 \\ \hline
       TF32 & FP32 & sp.m64n256k16 & 144.0/656.8 & 128.0/735.4 & 644.9 & 721.7 \\ \hline
       FP8 & FP16 & sp.m64n256k64 & 144.0/2619.9 & 128.0/2945.0 & 2588.6 & 2782.4 \\ \hline
       FP8 & FP32 & sp.m64n256k64 & 144.0/2622.8 & 128.0/2931.0 & 2588.7 & 2722.3 \\ \hline
       INT8 & INT32 & sp.m64n256k64 & 144.0/2612.4 & 128.0/2933.0 & 2593.9 & 2898.3 \\ \hline
      \end{tabularx}
      \vspace{-1.0 em}
      \end{table*}
    
\noindent\textbf{\emph{wgmma} results with different $N$ values.} We conduct tests using the example of wgmma.m64nNk16.f32.f16.f16, varying the value of $N$, and the results are presented in Table \ref{tab:Ndiff}. When $N$ is greater than or equal to 64, all \emph{wgmma} instructions can achieve throughputs that closely approach peak performance.
However, when $N$ is less than 64, the achieved throughputs decrease, and the ``SS'' mode of instructions exhibits higher latency than the ``RS'' mode, while the achieved throughputs are lower than those of the ``RS'' mode. As $N$ decreases, the computational density of \emph{wgmma} instructions gradually diminishes, making it challenging to conceal the latency associated with shared memory access, leading to the aforementioned phenomena.
Therefore, when utilizing wgmma instructions, it is advisable to opt for larger values of $N$ ($>=64$) whenever possible to attain superior performance.

\begin{table*}[]
\caption{Different \emph{wgmma} instructions with different $N$ values on H800 tensor cores. The definitions of LAT and throughput can be found in the caption of Table \ref{tab:densewgmma}. ``Rand" has the same latency as ``Zero".}
\label{tab:Ndiff}
\begin{tabular}{|l|llll|llll|}
\hline
\multicolumn{1}{|c|}{\multirow{2}{*}{\textbf{N}}} & \multicolumn{4}{c|}{\textbf{Dense}} & \multicolumn{4}{c|}{\textbf{Sparse}} \\ \cline{2-9} 
\multicolumn{1}{|c|}{} & \multicolumn{1}{l|}{\textbf{\begin{tabular}[c]{@{}l@{}}LAT/Throughput\\ (SS,Zero)\end{tabular}}} & \multicolumn{1}{l|}{\textbf{\begin{tabular}[c]{@{}l@{}}LAT/Throughput\\ (RS,Zero)\end{tabular}}} & \multicolumn{1}{l|}{\textbf{\begin{tabular}[c]{@{}l@{}}Throughput\\ (SS,Rand)\end{tabular}}} & \textbf{\begin{tabular}[c]{@{}l@{}}Throughput\\ (RS,Rand)\end{tabular}} & \multicolumn{1}{l|}{\textbf{\begin{tabular}[c]{@{}l@{}}LAT/Throughput\\ (SS,Zero)\end{tabular}}} & \multicolumn{1}{l|}{\textbf{\begin{tabular}[c]{@{}l@{}}LAT/Throughput\\ (RS,Zero)\end{tabular}}} & \multicolumn{1}{l|}{\textbf{\begin{tabular}[c]{@{}l@{}}Throughput\\ (SS,Rand)\end{tabular}}} & \textbf{\begin{tabular}[c]{@{}l@{}}Throughput\\ (RS,Rand)\end{tabular}} \\ \hline
256 & \multicolumn{1}{l|}{128.0/728.5} & \multicolumn{1}{l|}{128.0/731.9} & \multicolumn{1}{l|}{665.4} & 667.5 & \multicolumn{1}{l|}{144.0/1312.3} & \multicolumn{1}{l|}{128.0/1476.2} & \multicolumn{1}{l|}{1194.3} & 1277.5 \\ \hline
128 & \multicolumn{1}{l|}{64.0/728.5} & \multicolumn{1}{l|}{64.0/725.4} & \multicolumn{1}{l|}{659.8} & 661.7 & \multicolumn{1}{l|}{80.0/1176.4} & \multicolumn{1}{l|}{64.0/1463.3} & \multicolumn{1}{l|}{1109.6} & 1270.5 \\ \hline
64 & \multicolumn{1}{l|}{32.0/719.6} & \multicolumn{1}{l|}{32.0/719.7} & \multicolumn{1}{l|}{648.3} & 649.9 & \multicolumn{1}{l|}{48.0/977.4} & \multicolumn{1}{l|}{32.0/1450.1} & \multicolumn{1}{l|}{969.9} & 1263.4 \\ \hline
32 & \multicolumn{1}{l|}{24.0/477.3} & \multicolumn{1}{l|}{16.0/710.3} & \multicolumn{1}{l|}{471.5} & 634.4 & \multicolumn{1}{l|}{32.0/727.1} & \multicolumn{1}{l|}{18.0/1272.4} & \multicolumn{1}{l|}{723.4} & 1135.7 \\ \hline
16 & \multicolumn{1}{l|}{20.0/287.0} & \multicolumn{1}{l|}{13.0/434.2} & \multicolumn{1}{l|}{283.5} & 426.2 & \multicolumn{1}{l|}{24.0/482.3} & \multicolumn{1}{l|}{18.0/638.6} & \multicolumn{1}{l|}{479.8} & 636.3 \\ \hline
8 & \multicolumn{1}{l|}{18.0/158.2} & \multicolumn{1}{l|}{13.0/216.7} & \multicolumn{1}{l|}{157.6} & 215.2 & \multicolumn{1}{l|}{20.0/289.0} & \multicolumn{1}{l|}{16.0/359.4} & \multicolumn{1}{l|}{286.1} & 356.7 \\ \hline
\end{tabular}
\end{table*}

\noindent\textbf{Energy efficiency.} Although Tensor Cores have impressive performance, their energy efficiency should also be considered. Excellent energy consumption ratio will bring benefits both economically and environmentally. We choose mma for energy consumption testing because mma is the current compatible instruction, and previous codes can run directly on the latest hopper architecture. We test the largest operational shape in Table \ref{tab:syncwgmma}. The energy efficiency of \emph{mma} instructions is shown in Table \ref{tab:energy}. In terms of dense instructions, the average energy efficiency of H800 is 1.60 times and 1.69 times that of A100 and RTX4090 respectively. In terms of sparse instructions, the average energy efficiency of H800 is 1.33 times and 1.39 times that of A100 and RTX4090 respectively. We can find that the H800 has significantly higher energy efficiency.

\begin{table}[]
\caption{Power consumption and energy efficiency of maximum shape under mma instructions. T, D, and S represent Type, Dense, and Sparse respectively. P stands for energy, measured in Watts. E stands for efficiency, measured in TFLOPS/watt.}
\label{tab:energy}
\centering
\begin{tabular}{|l|l|c|cc|cc|cc|}
\hline
\multirow{2}{*}{\textbf{A/B}} & \multirow{2}{*}{\textbf{C/D}} & \multirow{2}{*}{\textbf{T}} & \multicolumn{2}{c|}{\textbf{A100}}           & \multicolumn{2}{c|}{\textbf{H800}}           & \multicolumn{2}{c|}{\textbf{4090}}           \\ \cline{4-9} 
                              &                               &                                & \multicolumn{1}{c|}{\textbf{P}} & \textbf{E} & \multicolumn{1}{c|}{\textbf{P}} & \textbf{E} & \multicolumn{1}{c|}{\textbf{P}} & \textbf{E} \\ \hline
\multirow{2}{*}{FP16}         & \multirow{2}{*}{FP16}         & D                              & \multicolumn{1}{c|}{173.4}      & 1.79       & \multicolumn{1}{c|}{188.6}      & 2.62       & \multicolumn{1}{c|}{189.1}      & 1.89       \\ \cline{3-9} 
                              &                               & S                              & \multicolumn{1}{c|}{198.8}      & 3.13       & \multicolumn{1}{c|}{187.2}      & 3.86       & \multicolumn{1}{c|}{214.0}      & 3.33       \\ \hline
\multirow{2}{*}{FP16}         & \multirow{2}{*}{FP32}         & D                              & \multicolumn{1}{c|}{188.5}      & 1.61       & \multicolumn{1}{c|}{196.7}      & 2.49       & \multicolumn{1}{c|}{154.1}      & 1.16       \\ \cline{3-9} 
                              &                               & S                              & \multicolumn{1}{c|}{216.1}      & 2.79       & \multicolumn{1}{c|}{194.9}      & 3.70       & \multicolumn{1}{c|}{165.9}      & 2.15       \\ \hline
\multirow{2}{*}{TF32}         & \multirow{2}{*}{FP32}         & D                              & \multicolumn{1}{c|}{214.7}      & 0.71       & \multicolumn{1}{c|}{254.9}      & 0.97       & \multicolumn{1}{c|}{174.3}      & 0.51       \\ \cline{3-9} 
                              &                               & S                              & \multicolumn{1}{c|}{235.7}      & 1.28       & \multicolumn{1}{c|}{232.5}      & 1.56       & \multicolumn{1}{c|}{187.9}      & 0.95       \\ \hline
\multirow{2}{*}{INT8}         & \multirow{2}{*}{INT32}        & D                              & \multicolumn{1}{c|}{178.4}      & 3.41       & \multicolumn{1}{c|}{165.3}      & 5.92       & \multicolumn{1}{c|}{201.4}      & 3.53       \\ \cline{3-9} 
                              &                               & S                              & \multicolumn{1}{c|}{193.9}      & 6.24       & \multicolumn{1}{c|}{163.3}      & 8.79       & \multicolumn{1}{c|}{219.8}      & 6.47       \\ \hline
\end{tabular}
\end{table}

\subsection{Transformer Engine Performance}
\begin{figure}[htbp]
        \centering
        \includegraphics[width=0.9\linewidth]{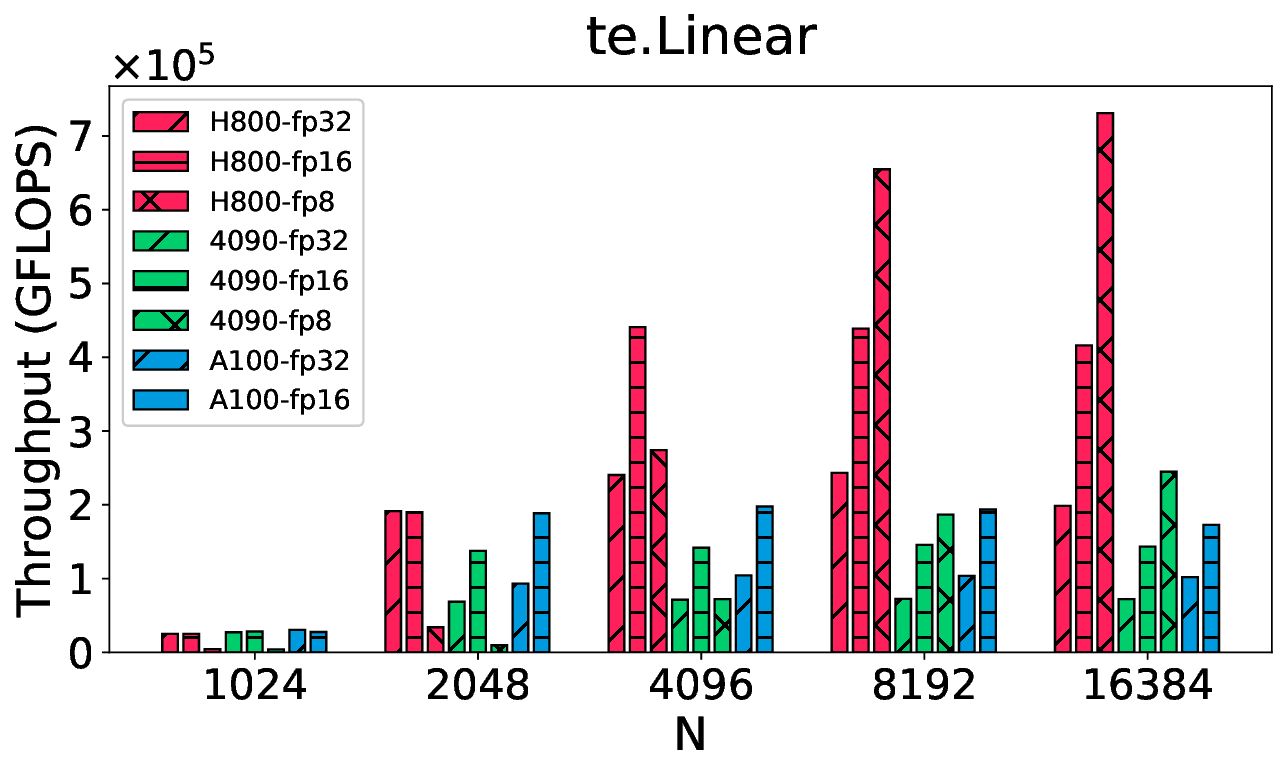} 
        \caption{Comparison of throughput for matrix multiplication with two same-size matrices $D(N*N) = A(N*N) \times B(N*N)$ in different hardware configurations and data types using \texttt{te.Linear}.}
        \label{fig:linear-all}
    \end{figure}

\noindent\textbf{\texttt{Te.Linear} analysis.} We extensively assess the Linear performance across diverse shapes, data types, and hardware setups (Fig. \ref{fig:linear-all}). Leveraging the Transformer Engine, we expedit matrix multiplications for the Linear layer using FP8 Tensor Cores. Our findings reveal an increase in GPU utilization and throughput with larger matrix sizes. FP8 performance is influenced by the overhead from data format conversion and quantization operators. For smaller matrix sizes, FP8 throughput is lower compared to FP16 or FP32. However, with $N$=8192, FP8's performance gains become evident. When $N$=16384, H800 and 4090 utilizing FP8 achieve almost twice the throughput of FP16. This underscores FP8's high throughput potential but underlines the need for specific conditions to attain optimal computing density.

\begin{figure}[htbp]
        \centering
        \includegraphics[width=0.95\linewidth]{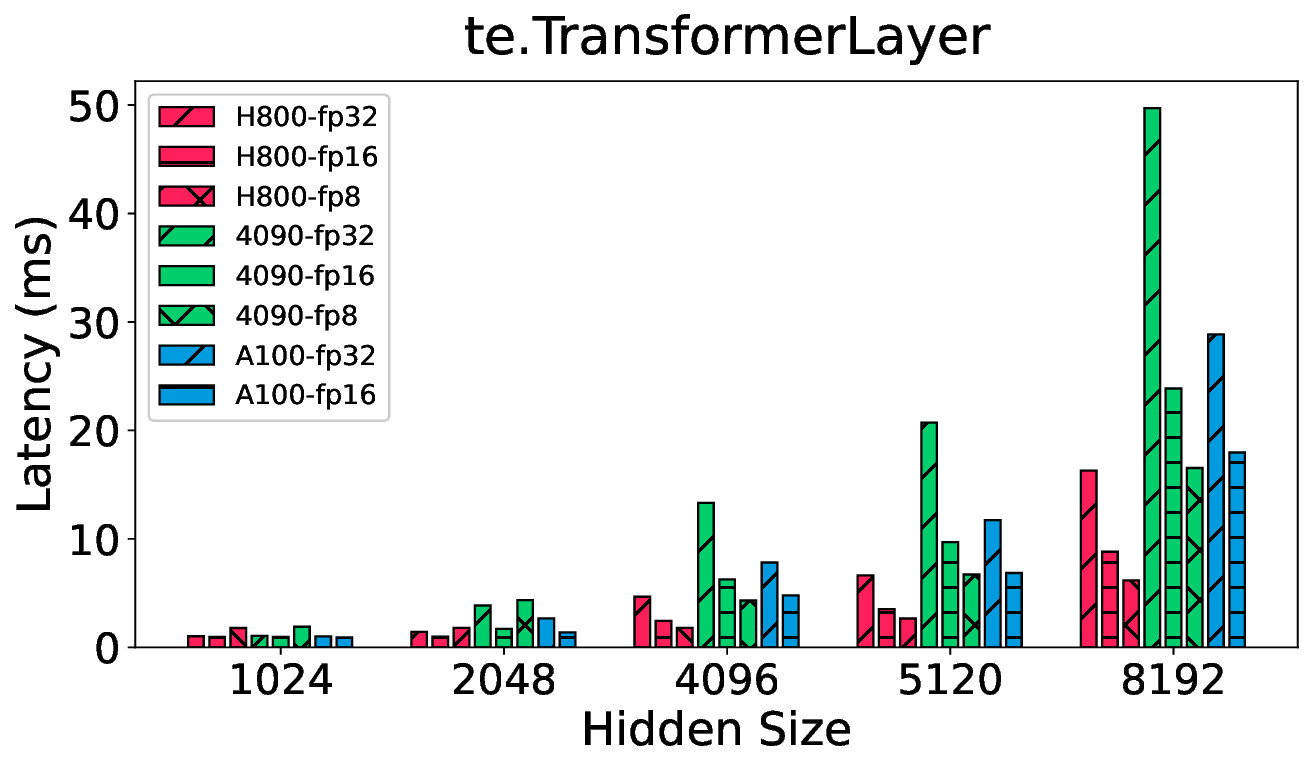} 
        \caption{Comparison of latency for the same input text in different hardware configurations and data types using \texttt{te.TransformerLayer}.}
        \label{fig:TeLayer-all}
        \vspace{-1.0 em}
    \end{figure}

\noindent\textbf{\texttt{Te.TransformerLayer} analysis.} 
The Transformer Engine condenses the entire Transformer Layer structure into \texttt{te.TransformerLayer}. Fig. \ref{fig:TeLayer-all} illustrates the latency for the same input text, offering a performance comparison across various hardware setups and data types with \texttt{te.TransformerLayer}. As computational density increases, the advantage of H800 in computation becomes evident. Notably, FP16 shows nearly twice the speed compared to FP32. FP8 outperforms FP16 for hidden\_size$>$4096 but does not achieve double FP16 performance. This is because some modules within the Transformer Layer still do not utilize FP8 precision for calculations and data movement.

    
\noindent\textbf{LLM Inference Throughput Results.}
        We test the state-of-the-art decode-only models on inference with different data types, as shown in Table \ref{tab:llm-inference}. We set the input length and text generation length to be relatively short, and the decode-only model is memory-bound during inference, so the computational advantages of FP8 Tensor Cores are not significant. Moreover, since the current Transformer Engine does not provide comprehensive support, data transmission between modules still occurs in FP16/FP32 without operator fusion. It is possible that when the model size and input data length increase, and with good operator fusion support, a certain improvement can be achieved.
    
    \begin{table}[!ht]
        \centering
        \caption{Inference Throughput (Tokens / s) for different model sizes on different GPUs and different data types}
        \label{tab:llm-inference}
        \begin{tabular}{|l|l|l|l|l|}
        \hline
            \textbf{GPU} & \textbf{Model} & \textbf{FP32} & \textbf{BF16} & \textbf{FP8} \\ \hline 
            \multirow{2}{*}{4090} & llama-3B & 414.08 & 425.19 & 429.31 \\ \cline{2-5}
            ~ & llama-2-7B & OOM & 350.69 & OOM \\ \hline
            \multirow{3}{*}{A100} & llama-3B & 674.50 & 670.87 & - \\ \cline{2-5}
            ~ & llama-2-7B & 400.88 & 548.57 & - \\ \cline{2-5} 
            ~ & llama-2-13B & OOM & 420.81 & - \\ \hline
            \multirow{3}{*}{H800} & llama-3B & 679.45 & 624.10 & 537.92 \\ \cline{2-5} 
            ~ & llama-2-7B & 568.91 & 502.65 & 474.42 \\ \cline{2-5}
            ~ & llama-2-13B & 357.57 & 399.38 & 356.11 \\ \hline
        \end{tabular}
    \end{table}




\subsection{New Features of Hopper}
\begin{figure*}[htb]
    \centering
    \includegraphics[width=0.98\linewidth]{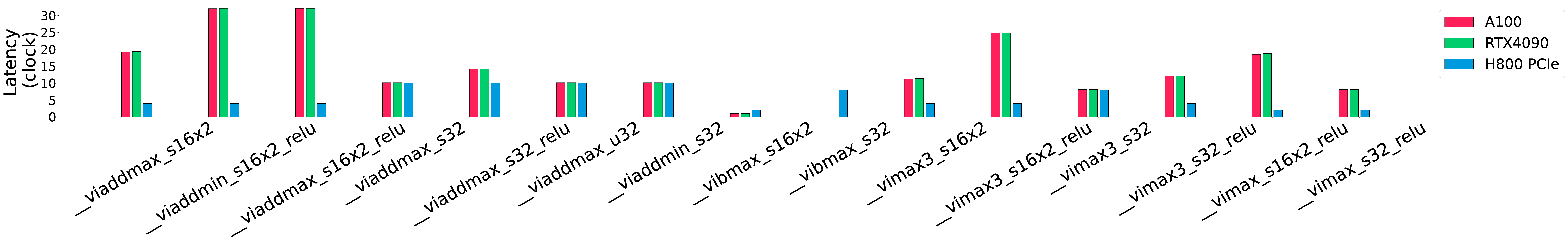} 
    \vspace{-1.2 em}
    \caption{Latency of DPX functions on different devices}
    \label{fig:dpx_lat}
    \vspace{-0.8 em}
\end{figure*}

\begin{figure*}[htb]
    \centering
    \includegraphics[width=0.98\linewidth]{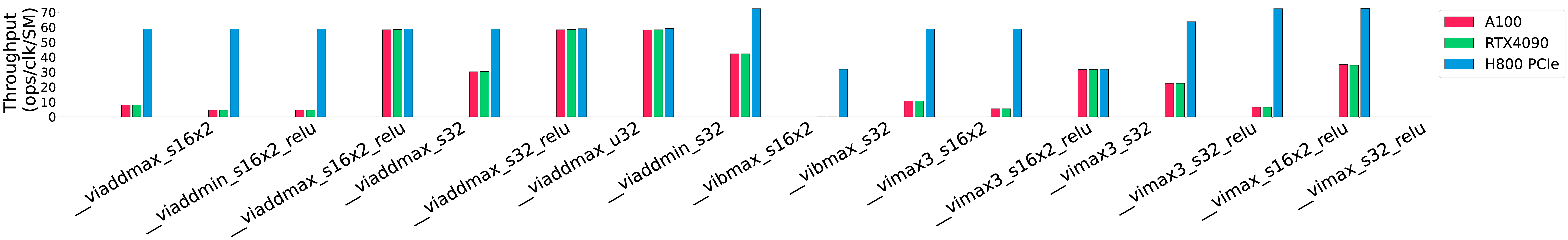} 
    \vspace{-1.2 em}
    \caption{Throughput of DPX functions on different devices}
    \label{fig:dpx_bw}
\end{figure*}

\noindent \textbf{DPX.} Fig. \ref{fig:dpx_lat} and Fig. \ref{fig:dpx_bw} show the latency and throughput of the DPX functions on three tested GPUs. Since the DPX of RTX4090 and A100 are software emulation, their performance is almost the same. What can be observed is that for \texttt{relu} instructions, the performance of H800 is significantly better than the other two. For 16-bit operations, H800 also has significant acceleration, up to 13 times.

However, not all functions have acceleration effects on Hopper. For some simple operations (e.g. \texttt{\_\_viaddmax\_s32}, which accepts 3 signed integers \texttt{(s1, s2, s3)} and returns \texttt{max(s1+s2,s3)}), we find that the performance of the three devices is close. In fact, by observing the SASS code, we find that new instructions (\texttt{VIMNMX}) were used on Hopper. Compared with previous \texttt{IMNMX}, performance does not seem to improve significantly. But in general, the Hopper architecture with DPX hardware acceleration has better performance than the previous generation architecture.

Additionally, \texttt{\_\_vibmax\_s32} data is not available on RTX4090 and A100. The reason is that compilation optimization optimizes this function into a max instruction. If we want to prevent this optimization, throughput measurements will be greatly affected. 

Another finding is that on H800, when the number of launched blocks is less than the number of SMs, the throughput of DPX functions is proportional to the number of blocks. When the number of blocks just exceeds an integral multiple of the number of SMs, the throughput plummets, gradually returning to the maximum level as the number of blocks increases. Maximum throughput occurs when the number of blocks is an integer multiple of the number of SMs. Therefore, we have enough reason to infer that the DPX acceleration unit is located at the SM level.

\noindent \textbf{Asynchronous Data Movement.}
Tables \ref{tab:async_h800} and \ref{tab:async_a100} illustrate the throughput comparison between \textsl{AsyncPipe} and \textsl{SyncShare} implementations on H800 and A100, respectively. Notably, \textsl{AsyncPipe} generally outperforms \textsl{SyncShare} with smaller block sizes (e.g., 8$\times$8 and 16$\times$16) on both GPUs. For instance, at a block size of 8$\times$8, \textsl{AsyncPipe} shows an average performance improvement of 39.5\% on H800 and 19.6\% on A100. The reason is that under small block sizes, insufficient warp numbers hinder hiding synchronous shared memory copy latency. On the contrary, the two-stage pipeline in \textsl{AsyncPipe} allows simultaneous data movement and computation across different stages.
However, as block size increases, the benefits diminish. Even with a block size of 32$\times$32 on H800, the throughputs of \textsl{AsyncPipe} are often worse than those of \textsl{SyncShare}. Larger block sizes result in high warp concurrency, effectively concealing shared memory copy latency.
\begin{table}[!ht]
\addtolength{\tabcolsep}{-3.5pt}
\caption{Benchmarking results of \textsl{globalToShmemAsyncCopy} on H800. The value of each cell represents the computational throughput measured in GFlops/s.}
\label{tab:async_h800}
    \centering
    \begin{tabular}{|c|c|c|c|c|c|c|c|}
    \hline
        \multicolumn{8}{|c|}{block size: 8$\times$8} \\ \hline
        Blocks/SM & 1 & 2 & 4 & 8 & 16 & 32 & Perf$\uparrow$ \\ \hline
        AsyncPipe & 516.69 & 998.45 & 1808.5 & 2931.29	& 3315.38 & 3615.99 & \multirow{2}{*}{39.5\%} \\
        SyncShare & 327.86 & 646.58 & 1191.48 & 2117.56 & 2736.06 & 2861.75 &  \\ \hline\hline
        \multicolumn{8}{|c|}{block size: 16$\times$16} \\ \hline
        Blocks/SM & 1 & 2 & 4 & 8 & 16 & 32 & Perf$\uparrow$ \\ \hline
        AsyncPipe & 2650.06 & 4531.02 & 5038.26 & 5510.76 & 5728.71 & 5929.61 & \multirow{2}{*}{9.7\%}\\ 
        SyncShare & 2372.41 & 3821.71 & 4713.84 & 5147.53 & 5309.23 & 5512.41 & \\ \hline\hline
        \multicolumn{8}{|c|}{block size: 32$\times$32} \\ \hline
        Blocks/SM & 1 & 2 & 4 & 8 & 16 & 32 & Perf$\uparrow$ \\ \hline
        AsyncPipe & 5570.17 & 6112.92 & 6372.73 & 6496.21 & 6592.66 & 6592.87 &\multirow{2}{*}{-1.8\%}\\
        SyncShare & 5782.03 & 6280.8 & 6465.53 & 6600.58 & 6649.46 & 6631.11 & \\ \hline
    \end{tabular}
\end{table}

\begin{table}[!ht]
\addtolength{\tabcolsep}{-3.5pt}
\caption{Benchmarking results of \textsl{globalToShmemAsyncCopy} on A100. The value of each cell represents the computational throughput measured in GFlops/s.}
\label{tab:async_a100}
    \centering
    \begin{tabular}{|c|c|c|c|c|c|c|c|}
    \hline
        \multicolumn{8}{|c|}{block size: 8$\times$8} \\ \hline
        Blocks/SM & 1 & 2 & 4 & 8 & 16 & 32 & Perf$\uparrow$ \\ \hline
        AsyncPipe & 379.03 & 798.5 & 1544.15 & 2429.93 & 2825.64 & 2888.84 & \multirow{2}{*}{19.6\%} \\
        SyncShare & 379.03 & 742.93 & 1325.88 & 1982.38 & 2112.6 & 2256.17 &  \\ \hline\hline
        \multicolumn{8}{|c|}{block size: 16$\times$16} \\ \hline
        Blocks/SM & 1 & 2 & 4 & 8 & 16 & 32 & Perf$\uparrow$ \\ \hline
        AsyncPipe & 2198.21 & 2566.83 & 3821.09 & 4205.72 & 4413.69 & 4527.82 & \multirow{2}{*}{4.9\%}\\ 
        SyncShare & 1754.73 & 2974.9 & 3724.42 & 4015.96 & 4207.57 & 4316.63 & \\ \hline\hline
        \multicolumn{8}{|c|}{block size: 32$\times$32} \\ \hline
        Blocks/SM & 1 & 2 & 4 & 8 & 16 & 32 & Perf$\uparrow$ \\ \hline
        AsyncPipe & 4453.52 & 4863.73 & 5020.21 & 5106.74 & 5150.78 & 5129.68 &\multirow{2}{*}{1.7\%}\\
        SyncShare & 4428.55 & 4917.25 & 5024.77 & 5025.45 & 4996.66 & 5028.47 & \\ \hline
    \end{tabular}
\end{table}

\noindent \textbf{Distributed Shared Memory.}
SM-to-SM network latency is 180 cycles, a 32\% reduction compared to L2 cache. This validates the advantages of the network, facilitating efficient data exchange from producers to consumers.
\begin{figure}[htbp]
    \centering
    \includegraphics[width=1.0\linewidth]{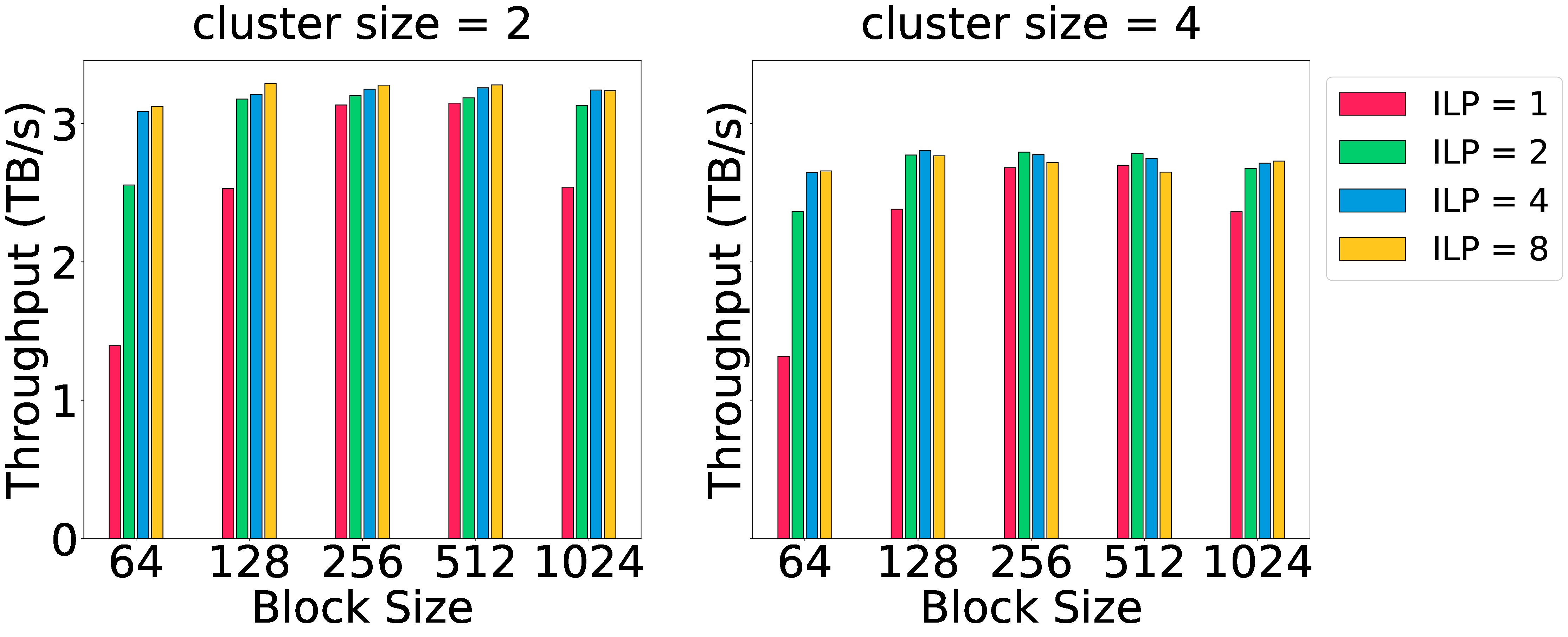}  
    \caption{The data communication throughput of the SM-to-SM network. ``ILP'' refers to the number of parallelizable data movement instructions.}
    \label{fig:dsm_throughput}
\end{figure}

\begin{figure}[htbp]
    \centering
    \includegraphics[width=1.0\linewidth]{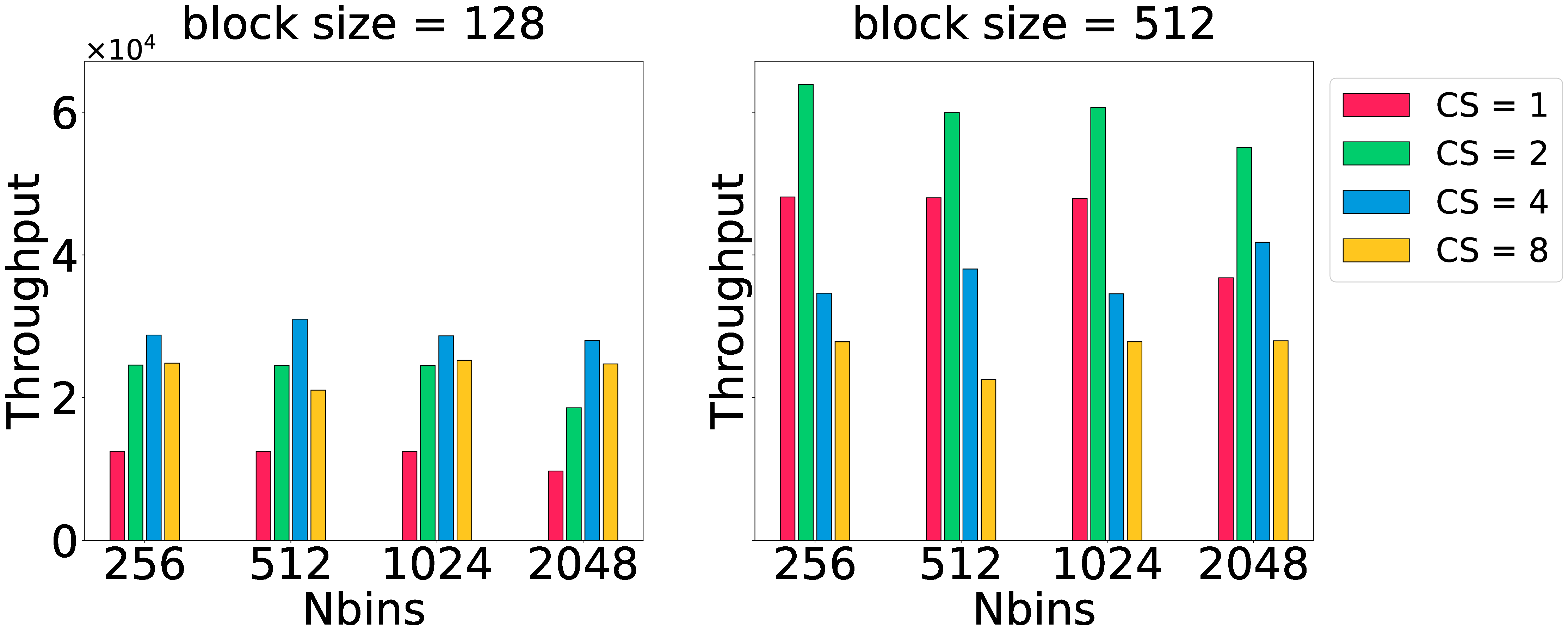}  
    \caption{Performance of the histogram application with distributed shared memory. The throughput is measured by the number of processing elements per second. ``CS'' refers to the cluster size. ``Nbins'' refers to the number of histogram bins.}
    \label{fig:histogram}
\end{figure}

In Fig. \ref{fig:dsm_throughput}, SM-to-SM throughput is illustrated for varying cluster and block sizes. As typically observed in similar benchmarks, larger block sizes and more parallelizable instructions result in higher throughputs. A peak throughput of nearly 3.27 TB/s is observed with a cluster size of 2, reducing to 2.65 TB/s with a cluster size of 4. Interestingly, as more blocks in the cluster compete for SM-to-SM bandwidth, the overall throughput gets lower and lower. While a larger cluster size can reduce data movement latency for more blocks, it intensifies throughput competition. Balancing this tradeoff by selecting optimal block and cluster sizes is an important direction for exploration.

Fig. \ref{fig:histogram} displays the histogram throughput with distributed shared memory. 
First, the optimal cluster size differs for various block sizes (CS=4 for block size 128, CS=2 for block size 512). Increasing block and cluster sizes can saturate SM-to-SM network utilization, potentially degrading overall performance due to resource contention.
Second, a notable performance drop occurs from 1024 to 2048 \textsl{Nbins} when CS=1. Larger \textsl{Nbins} demand more shared memory space and limit active block numbers on an SM. Employing the cluster mechanism to divide \textsl{Nbins} within the same cluster enhances block concurrency, mitigating this issue.
Lastly, although shared memory is not a limiting factor for active block numbers with block size = 512, choosing an appropriate cluster size ease the on-chip shared memory traffic by leveraging the SM-to-SM network resource, ultimately improving overall performance.

\section{Conclusion}

This paper delves into memory hierarchy and tensor core performance of the newest three Nvidia GPU architectures using instruction-level benchmarks. We found that the hopper architecture shows advantages in both memory bandwidth and tensor core that are consistent with official claims. It is worth noting that on tensor core, we need to use the latest wgmma instructions to take advantage of all the performance of the fourth generation tensor core. We analyze AI performance across diverse architectures at library and application levels, emphasizing the impact of varied precisions. Experiments show that when the operation scale is relatively large, low-precision data types will show greater advantages. Additionally, we explore key features of the Hopper architecture: DPX, asynchronous data movement, and distributed shared memory. Our research enhances comprehension of the latest architecture's traits and performance, aiding in optimized algorithm design and application performance.



\section*{Acknowledgments}
This research was also supported by the National Natural Science Foundation of China (No. 62302126), the Shenzhen Science and Technology Program (No. RCBS20221008093125065, No. JCYJ20220818102414030, No. JSGGKQTD20221101115655027), a Hong Kong RGC RIF grant under the contract R6021-20, and Hong Kong RGC CRF grants under the contracts C7004-22G and C1029-22G.

\bibliographystyle{IEEEtran}
\bibliography{microbenchmarks.bbl}

\end{document}